\documentclass[twocolumn,aps,showpacs]{revtex4-1}
\usepackage{amsfonts}
\usepackage{amsmath}
\usepackage{graphicx}
\usepackage{bm}
\usepackage{amssymb}
\usepackage{dcolumn}
\usepackage{color}

\begin{document}

\title{Theory of the   spin  Hall  effect, and its inverse,  in a ferromagnetic metal near the Curie temperature}

\author{B. Gu$^{1,2}$,
T. Ziman$^{3,4}$, and S. Maekawa$^{1,2}$}

\affiliation{$^1$Advanced Science Research Center, Japan Atomic Energy Agency, Tokai 319-1195, Japan
\\ $^{2}$CREST, Japan Science and Technology Agency, Sanbancho, Tokyo 102-0075, Japan
\\ $^{3}$ Institut Laue Langevin, Bo\^\i te Postale 156, F-38042 Grenoble Cedex 9, France 
\\ $^{4}$ LPMMC  (UMR 5493), Universit\'e Grenoble 1 and CNRS, 38042 Grenoble, France}

\begin{abstract}
We give a theory of the inverse spin Hall effect (ISHE) in ferromagnetic metals based on skew scattering 
via collective spin fluctuations. This extends Kondo's theory of the anomalous Hall effect (AHE) to include short-range 
spin-spin correlations. We find a relation between the ISHE and the four-spin correlations 
near the Curie temperature $T_{\rm C}$. Such four-spin correlations do not contribute to the AHE, which relates to the three-spin correlations.
Thus our theory shows an essential difference between the AHE and ISHE,
providing an essential complement to Kondo's  classic theory of the AHE in metals. 
We note the relation to skew-scattering mechanisms based 
on impurity scattering. Our theory can be compared to recent experimental results
by Wei {\it et al.} [Nat. Commun. {\bf 3}, 1058 (2012)] for the ISHE in ferromagnetic alloys.     
\end{abstract}

\pacs{} \maketitle

The spin Hall effect (SHE), which converts charge
current into spin current, along with the inverse effect (ISHE) which re-converts the currents, is one
of the key phenomena for the further development of
spintronic devices \cite{Dyakonov, SHE-Hirsch}. A difficulty in exploiting the effect is that it depends on a spin-orbit interaction  which, as a relativistic effect, is intrinsically weak except in heavy elements.
Recently the doping of gold with low densities of  iron impurities was
proposed as a mechanism to enhance the SHE through spin fluctuations of the individual iron
moments \cite{AuFe-Guo,AuFe-Gu}. This raises the question as to whether the effect might also be sensitive to {\it collective} magnetic
fluctuations, in particular, close to a magnetic phase transition, where the fluctuations
are strong, in analogy with 
the anomalous Hall effect (AHE) in a pure ferromagnetic metal
such as Fe \cite{AHEFe} and Ni \cite{AHENi}, where a  peak of the Hall resistivity appears
below the Curie temperature $T_{\rm C}$.  Wei {\it et al.} \cite{Otani} observed a feature in the ISHE, scaling with $T_{\rm_C}$ for different samples of a metallic alloy, which suggested that this is in fact so, but the form is quite different from that of the AHE.

The  observations of the AHE near the Curie temperature were explained by Kondo in his calculations \cite{Kondo} of an $s$-$d$ (or $s$-$f$) lattice 
where, apart from the standard Coulomb interactions between the orbitals, 
he included on-site  spin-orbit interactions. This gives rise to  skew scattering in the conducting $s$ band. 
We will show here that while both the AHE and SHE are generated by skew scattering that is linear in the spin-orbit term, they have {\it qualitatively} different behaviors near $T_{\rm C}$  as they depend on correlations of distinct order.  
In Kondo's formulation  Coulomb interactions are included by keeping $s$- and $p$-scattering channels for the conducting electrons around each localized orbital.  The angular momentum can thereby be transferred from the local moment to the conduction electrons.
As we are interested in the effects of skew scattering, we will write only terms depending on the vector product $\pmb{\kappa}^{\prime}\times\pmb{\kappa}$, where  $\pmb{\kappa}$ (respectively $\pmb{\kappa}^{\prime}$) is the unit vector in the direction of $\bf{k}$ (respectively $\bf{k}^{\prime}$), the momentum of an electron in the $s$ band. The $s$-$d$ exchange interaction is then \cite{Kondo} (omitting terms depending on quadrupole moments that do not contribute here) 
\begin{equation}
\begin{split}
H&=-\sum_{n,\textbf{k},\textbf{k}^{\prime},\mu,\mu^{\prime}}N^{-1}e^{i(\textbf{k}^{\prime}-\textbf{k})\cdot\textbf{R}_n}
a^{\ast}_{\textbf{k}\mu}a_{\textbf{k}^{\prime}\mu^{\prime}}
\Bigl [ (3i/4)F_2\pmb{L}_n\cdot(\pmb{\kappa}^{\prime}\times\pmb{\kappa}) \\
+&2(\pmb{S}_n\cdot\pmb{s}_c)\bigl\{J(\textbf{k},\textbf{k}^{\prime})+(3i/2)c_2F_2\pmb{L}_n\cdot(\pmb{\kappa}^{\prime}\times\pmb{\kappa})\bigr\}\Bigr ].
\end{split} \nonumber
\end{equation}
 $\pmb{L}_n$ and  $\pmb{S}_n$ are  the  total orbital and spin angular momenta of the localized $d$ orbital on atomic  site $n$ at position $\textbf{R}_n$, of a total number of 
$N$ magnetic atoms in the crystal. $\pmb{s}_c$ is the spin operator for the conduction electrons. $J(\textbf{k},\textbf{k}^{\prime})=F_0+2F_1(\pmb{\kappa}\cdot\pmb{\kappa}^{\prime})$ with
 $F_{0,1,2}$   exchange terms generated by Coulomb interactions between the localized $d$ electrons 
and the different channels of scattering with propagating $s$ orbitals. 
The coefficients $c_2$ appear in the summation of individual orbital and spin angular momenta of each electron within the localized orbitals 
to give the total  momentum \cite{Kondo}.
Assuming an orbital singlet but adding a small spin-orbit $H_{LS} = \lambda\sum_{n}\pmb{L}_n\cdot\pmb{S}_n$, 
it generates operators $\pmb{L}_n$ on each site and gives a corrected Hamiltonian with the coefficient $\Lambda_1$ linear in the  spin-orbit coupling $\lambda$ and determined by the local crystal field levels \cite{Kondo},
\begin{equation}
\begin{split}
&H =-\sum_{n,\textbf{k},\textbf{k}^{\prime},\mu,\mu^{\prime}}N^{-1}e^{i(\textbf{k}^{\prime}-\textbf{k})\cdot\textbf{R}_n}
a^{\ast}_{\textbf{k}\mu}a_{\textbf{k}^{\prime}\mu^{\prime}}
\Bigl [ 2(\pmb{S}_n\cdot\pmb{s}_c)J(\textbf{k},\textbf{k}^{\prime}) \\
&
+i\Lambda_1F_2 
\bigl\{\pmb{S}_n\cdot(\pmb{\kappa}^{\prime}\times\pmb{\kappa})
+2c_2\left(\pmb{S}_n\cdot\pmb{s}_c\right)\left(\pmb{S}_n\cdot\left(\pmb{\kappa}^{\prime}\times\pmb{\kappa}\right)\right) \\
&+2c_2\left(\pmb{S}_n\cdot\left(\pmb{\kappa}^{\prime}\times\pmb{\kappa}\right)\right)\left(\pmb{S}_n\cdot\pmb{s}_c\right)
-4c_2/3\left(\pmb{S}_n\cdot\pmb{S}_n\right)
\left(\pmb{s}_c\cdot(\pmb{\kappa}^{\prime}\times\pmb{\kappa})\right)\bigr\}\Bigr ].
\end{split}
\label{E-Ham}
\end{equation}
Considering only elastic scattering, the matrix element of the above Hamiltonian is \cite{Kondo},
\begin{equation}
\begin{split}
&H_{\textbf{k}\pm,\textbf{k}^{\prime}\pm}
=-\sum_{n}N^{-1}e^{i(\textbf{k}^{\prime}-\textbf{k})\cdot\textbf{R}_n}
\Bigl [ \pm J(\textbf{k},\textbf{k}^{\prime}) (M_n-\langle M_n\rangle) \\
&+i\Lambda_1F_2(\pmb{\kappa}^{\prime}\times\pmb{\kappa})_{\pmb{\zeta}}
\bigl\{ \pm 2c_2(M_n^2-\langle M_n^2\rangle)+( M_n-\langle M_n \rangle)\bigr\}\Bigr ].
\end{split}
\label{E-Hkk}
\end{equation}
 $M_n$ is the magnetization of the localized electron of the $n$-th ion,
and $\pmb{\zeta}$ is the magnetization direction below $T_{\rm C}$ or the direction of polarization of the spin current above, a direction 
which will be determined by the spin injector in the devices used. Subtractions of the lattice-averaged powers of the magnetization  at each site, $\langle M_n \rangle $
and $\langle M_n^2 \rangle$,  appear because any modification of the crystalline potential simply renormalizes the Bloch functions
of the conduction electrons and does not contribute to scattering.
There are two parts in Eq. (\ref{E-Hkk}): One is the $J(\textbf{k},\textbf{k}^{\prime})$ term,
coming from the $s$-$d$ scattering,
and the other is the $\Lambda_1$ term, which is linear in spin-orbit coupling $\lambda$.
$\pm$ are spin states of the conduction electron.
$(\pmb{\kappa}^{\prime}\times\pmb{\kappa})_{\pmb{\zeta}}$ comes from the
approximation of including elastic scattering only.

In order to find a net current transverse to the voltage, the transition probabilities must be calculated to at least the second Born approximation , {\it i.e.,} the transition probabilities are to third power in the Hamiltonian.
There is a triple summation over sites which are reduced, with Kondo's assumption of uncorrelated fluctuations, to a single summation. Now we generalize to include correlations but for simplicity we include those with only two-site indices, but all powers of spin. 
The transition probability
from $\textbf{k}^{\prime}\pm$ to $\textbf{k}\pm$ is given by
\begin{equation}
W(\textbf{k}^{\prime}\pm,\textbf{k}\pm)=\delta(E_k-E_{k^{\prime}})
\bigl\{U_1+U_3+V^{\pm}(\textbf{k}^{\prime},\textbf{k})\bigr\},
\nonumber
\end{equation}
where the $s$-$d$ scattering contribution of $U_1$ is given by
\begin{eqnarray}
&&U_1=(2\pi/\hbar N)|J(\textbf{k},\textbf{k}^{\prime})|^2p_1^{},
\nonumber
\\ 
&&p_1^{}=\sum^{N-1}_{n=0}e^{i(\textbf{k}^{\prime}-\textbf{k})\cdot(\textbf{R}_n-\textbf{R}_0)}
\langle(M_n-\langle M_n\rangle)(M_0-\langle M_0\rangle)\rangle. \nonumber
\end{eqnarray}
A contribution $U_3=(2\pi/\hbar N)P(\textbf{k},\textbf{k}^{\prime})$  is included to represent non-magnetic contributions, static impurity scattering, or phonons,  
where $P(\textbf{k},\textbf{k}^{\prime})$ is assumed to depend only on the 
angle between $\textbf{k}$ and $\textbf{k}^{\prime}$, and not to show significant temperature dependence close to the magnetic phase transition.
The spin-orbit contribution of $V^{\pm}(\textbf{k}^{\prime},\textbf{k})$ is obtained as 
\begin{equation}
\begin{split}
&V^{\pm}(\textbf{k}^{\prime},\textbf{k})
=(2\pi/\hbar)(\Lambda_1F_2V/3\pi N^2)(2m^3)^{1/2}\hbar^{-3}E_{k}^{1/2}\\
&\times (\pmb{\kappa}^{\prime}\times\pmb{\kappa})_{\pmb{\zeta}}
\Bigl [( r_1^{} \pm r_{2a}^{})
\bigl\{3F_{0}^{2}+4F_1^2(\pmb{\kappa}\cdot\pmb{\kappa}^{\prime})\bigr\}
+( r_1^{} \pm r_{2b}^{})\\
&\times \bigl\{-4F_{0}F_{1}-8F_1^2(\pmb{\kappa}\cdot\pmb{\kappa}^{\prime})\bigr\}
\Bigr ],
\end{split}
\label{E-a2-V}
\end{equation}
where $r_1^{}$, $r_{2a}^{}$, and $r_{2b}^{}$ depend on $\textbf{k}^{\prime}-\textbf{k}$ and temperature, 
\begin{equation}
r_1^{} = \sum^{N-1}_{n=0}e^{i(\textbf{k}^{\prime}-\textbf{k})\cdot(\textbf{R}_n-\textbf{R}_0)}
\langle (M_n-\langle M_n\rangle)^2(M_0-\langle M_0\rangle)\rangle, \nonumber
\end{equation}
\begin{equation}
r_{2a}^{} = 2c_2\sum^{N-1}_{n=0}e^{i(\textbf{k}^{\prime}-\textbf{k})\cdot(\textbf{R}_n-\textbf{R}_0)}
\langle (M_n-\langle M_n\rangle)^2(M_0^2-\langle M_0^2\rangle)\rangle, \nonumber
\end{equation}
\begin{equation}
\begin{split}
r_{2b}^{} = 2c_2\sum^{N-1}_{n=0}e^{i(\textbf{k}^{\prime}-\textbf{k})\cdot(\textbf{R}_n-\textbf{R}_0)}
\langle (M_n-\langle M_n\rangle)(M_n^2-\langle M_n^2\rangle) \\
\times(M_0-\langle M_0\rangle)\rangle. \nonumber
\end{split}
\end{equation}   
These generalize the on-site $n$ = 0 terms, which we denote as $r_1^{\textrm{o}}$ or $r_2^{\textrm{o}}$, of Kondo.  There are now two distinct contributions at fourth order in spin, $r_{2a}^{}$ and $r_{2b}^{}$. All off-site  coefficients ($n\ne 0$)  depend on the momentum transfer, making integrals  slightly more complex than Kondo's.

\begin{figure}[tbp]
\includegraphics[width = 6.5 cm]{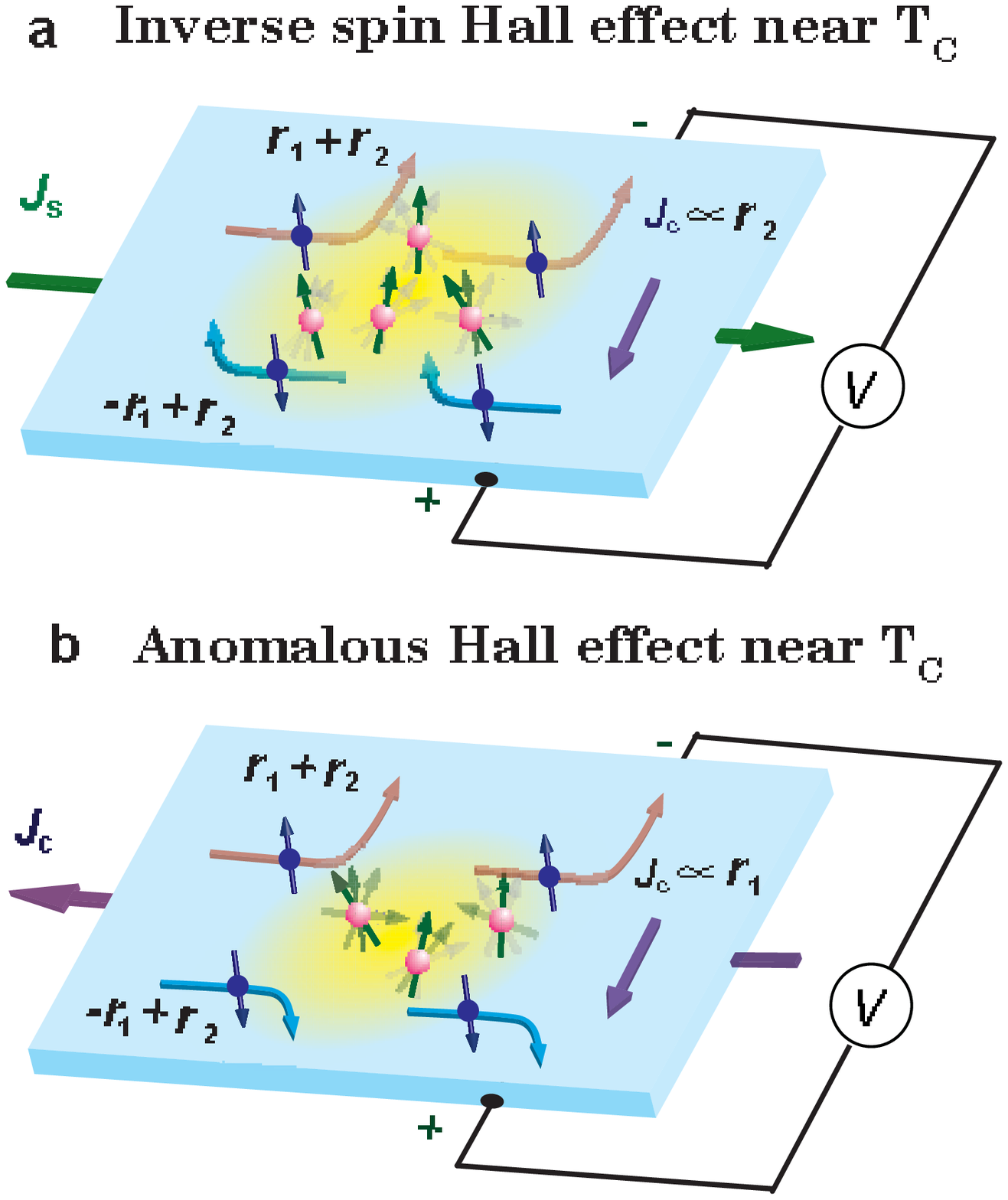}
\caption{(Color online) Schematic picture of (a) ISHE and (b) AHE near $T_{\rm C}$. The skew scattering amplitudes for spin-up and spin-down conduction electrons (represented by arrows on the deflecting arrows) are proportional to $r_1 + r_2$ and $-r_1+ r_2$, respectively. The potentials measured depend on correlations of 4 (respectively  3),  local spins in (a)  [respectively  (b)].}
\label{FigAHESHE}
\end{figure}

\emph{Inverse spin Hall effect}. Here there are two spin polarizations possible:  first,  that of the spin current,  introduced  by a spin injector, assumed to have a much higher Curie temperature than the $T_{\rm C}$ of the ferromagnetic metal (spin detector) in which 
the skew scattering and the inverse spin Hall effect occur. 
It is the injector which defines the direction of quantization of the spin current $\pmb{\zeta}$  both above and below $T_{\rm C}$.  
The second is the ordered moment of the ferromagnetic metal (spin detector), which we assume to be parallel to the spin injector below $T_{\rm C}$.
The incident spin currents are driven by the spin diffusion field: 
the difference in gradients of two electrochemical potentials $\varepsilon^{\pm}_F$ \cite{Takahashi,Takahashi2},
$\pmb{\mathcal{F}}= -\frac{1}{e}\nabla (\varepsilon_{F}\pm\delta\varepsilon_{F}) $ 
with magnitude $\mathcal{F}$ and direction $\pm\pmb{\sigma}$ for the spin states $\pm$.
The spin current is scattered to create a transverse voltage. 
Solving the Boltzmann equations for distributions  over the two Fermi surfaces of spin-up and  spin-down electrons using 
the skew scattering probabilities Eq. (\ref{E-a2-V}), the current densities \cite{Kondo} can be calculated as 
\begin{equation}
\textbf{j}^{\pm}=(\sigma_{\|}/2)\mathcal{F}
\Bigl\{\pm(\pmb{\sigma})\pm (\Psi_{\pm}/\Phi)
\bigl(\pmb{\zeta}\times\pmb{\sigma}\bigr)\Bigr\},\nonumber
\end{equation}
where 
\begin{eqnarray}
&&\frac{\Psi_{\pm}}{\Phi}=(\Lambda_1F_2V/3\pi N)(2m^3)^{1/2}\hbar^{-3}E_{k}^{1/2}\frac{B_{\pm}}{A^{{\textrm{}}}}, \label{E-psipmphi}\\
&&B_{\pm} = \frac{1}{8\pi}  \int_0^{\pi}\int_0^{2\pi}
\bigl\{ (r_1^{}\pm r_{2a}^{})(3F_0^2+4F_1^2\cos\theta)\notag\\
&&+(r_1^{}\pm r_{2b}^{})(-4F_0F_1-8F_1^2\cos\theta) \bigr\}\sin^3\theta d\theta d\phi. 
\nonumber 
\end{eqnarray}
and
\begin{eqnarray}
\begin{split}
A^{}=\frac{1}{4\pi} & \int_0^{\pi}\int_0^{2\pi}
\bigl\{|J(\textbf{k},\textbf{k}^{\prime})|^2p_1^{} (\textbf{k}^{\prime}-\textbf{k} ) \\
& +P(\textbf{k},\textbf{k}^{\prime})\bigr\}(1-\cos\theta)\sin\theta  d\theta d\phi.
\nonumber 
\end{split}
\end{eqnarray}
$p_1^{} $  depends on all components of momentum transfer as it includes 
spin-spin correlations between different ions. $\theta$ and $\phi$ are  the angles for  spherical coordinates  with respect to the  longitudinal current.
$\sigma_{\|}^{-1}$ is the resistivity given in terms of  the integral $A^{{\textrm{}}}$,
\begin{equation}
\sigma_{\|}^{-1}=\rho_{\|}=(3\pi m/2\hbar e^2)(V/N)(A^{}/E_F).
\label{E-a2-sigma0}
\end{equation}
Thus, the spin current $\textbf{j}^s\equiv \textbf{j}^{+} - \textbf{j}^{-}$
and the charge current $\textbf{j}^c \equiv \textbf{j}^{+} + \textbf{j}^{-}$ are
\begin{eqnarray}
\textbf{j}^{s} &=&\sigma_{\|}\mathcal{F}\pmb{\sigma}
+\sigma_{\|}\mathcal{F}\bigl\{(\Psi_{+}+\Psi_{-})/(2\Phi)\bigr\}
(\pmb{\zeta}\times\pmb{\sigma}), \label{E-jsshe}\\
\textbf{j}^{c} &=&\sigma_{\|}\mathcal{F}\bigl\{(\Psi_{+}-\Psi_{-})/(2\Phi)\bigr\}
(\pmb{\zeta}\times\pmb{\sigma}). 
\label{E-jcshe}
\end{eqnarray}   
As schematically shown in Fig. \ref{FigAHESHE}(a), for the incident spin current in the ISHE configuration, the spin-up ($+$) and spin-down ($-$) conduction electrons are scattered by skew scattering to the same side with the transition probability proportional to the terms $r_1 + r_2$ ($r_2$ represents $r_{2a}$ and $r_{2b}$) and $-r_1 + r_2$, respectively. As a result, the Hall current $\pmb{j}^c$ is proportional to $r_2$.

The Hall resistivity is defined as $\rho_{\textrm{ISH}}=E_{\bot}/j_{\|}$.
It has $j_{\|}=\sigma_{\|}\mathcal{F}$ from Eq. (\ref{E-jsshe}),
and $E_{\bot}=\mathcal{F}(\Psi_{+}-\Psi_{-})/(2\Phi)$ from Eq. (\ref{E-jcshe}). 
In the direction $\pmb{\zeta}\times\pmb{\sigma}$,
the spin current does not contribute to the electric field $E_{\bot}$.
Combining Eqs. (\ref{E-psipmphi}) and (\ref{E-a2-sigma0}),  the inverse spin Hall resistivity $\rho_{\textrm{ISH}}$ can be obtained as 
\begin{eqnarray}
\rho_{\textrm{ISH}} = \Lambda_1B_2F_2
(V/N)^2(2m^5)^{1/2}/(2\hbar^{4}e^2E_{F}^{1/2}), \label{E-rhoshe} 
\end{eqnarray}
where $B_2=(B_{+}-B_{-})/2$. 
 $B_2$ involves only $r_{2a}^{}$ and $r_{2b}^{}$:
\begin{eqnarray}
&&B_2 = \frac{1}{8\pi}\int_0^{\pi}\int_0^{2\pi}
\bigl\{ r_{2a}^{}(3F_0^2+4F_1^2\cos\theta)\notag\\
&&+r_{2b}^{}(-4F_0F_1-8F_1^2\cos\theta) \bigr\}\sin^3\theta d\theta d\phi. \nonumber
\end{eqnarray}
We find that the 
temperature variation of the Hall resistivity $\rho_{\textrm{ISH}}$ for the inverse spin  Hall effect in a ferromagnetic metal
is that of  $B_2$, {\it i.e.}, fourth-order correlations. 
The other factors in Eq. (\ref{E-rhoshe}) do not show a significant temperature dependence close to the magnetic phase transition \cite{Kondo}.
Note the on-site term  $r_{2a}^{\textrm{o}}=r_{2b}^{\textrm{o}}$ is independent of angles 
$\theta$ and $\phi$ so the integral gives, in a  purely local approximation, $\rho_{\textrm{ISH}} ^{\textrm{o}} \propto 
 \Lambda_1r_2^{\textrm{o}}(F_0^2-4F_0F_1/3)F_2$. This  is analogous to the local approximation $\rho_{xy}$ for the anomalous Hall resistance obtained by Kondo, but appears to compare less well with experiment \cite{Otani}: For the ISHE the non-local terms are more significant.

\emph{Spin Hall effect}. For this effect \cite{Takahashi}  a transverse spin current is generated from a charge current, so we can deduce the current equations from those of the inverse effect by interchanging the spin and charge labels
of the currents and 
replacing the spin diffusion field $\mathcal{F}$  by  the driving external electric field $\mathcal{E}$. 
To define a spin Hall transport coefficient we must specify the geometry of the measurement.  We consider the non-local device scheme \cite{Takahashi} where the potentials are measured for parallel ($V^{P}_{\bot}$) and antiparallel ($V^{AP}_{\bot}$)  magnetizations of the    injecting and detecting ferromagnetic electrodes.
Both  $V^{P}_{\bot}$ and $V^{AP}_{\bot}$  are proportional to  $ \mathcal{E}(\Psi_{+}-\Psi_{-})/(2\Phi)$ with different coefficients of proportionality. 
$\Psi_{+}$ and $\Psi_{-}$ follow from  Eqs. (\ref{E-psipmphi}) and (\ref{E-a2-sigma0}) as before,  
so the spin accumulation signal $\rho_{\textrm{SH}}=(V^{P}_{\bot}-V^{AP}_{\bot})/j_{\|}\propto \rho_{\textrm{ISH}}$.    
The proportionality constant  depends on the transparencies of the barriers,  which should be essentially independent of temperature, so that the temperature variation of the spin  Hall resistivity $\rho_{\textrm{SH}}$
is just  that of the inverse spin Hall coefficient $\rho_{\textrm{SH}}$, {\it i.e.}, $B_2$.  
 
\emph{Anomalous Hall effect}. We also generalized the expression for the AHE to include short-range spin-spin correlations. 
By interchanging the spin and charge labels
of the currents and 
replacing the spin diffusion field $\pmb{\mathcal{F}}$  by  the external electric field $\pmb{\mathcal{E}}=-\nabla\phi$
with magnitude $\mathcal{E}$ and direction $\pmb{\sigma}$ for the spin states $\pm$,
the current densities \cite{Kondo} can be calculated as
\begin{equation}
\textbf{j}^{\pm}=(\sigma_{\|}/2)\mathcal{E}
\Bigl[\pmb{\sigma}+(\Psi_{\pm}/\Phi)
\bigl\{\pmb{\zeta}\times\pmb{\sigma}\bigr\}\Bigr]. 
\nonumber 
\end{equation}
Thus, the spin 
and  charge currents  are
obtained as
\begin{eqnarray}
\textbf{j}^{s} &=& \sigma_{\|}\mathcal{E}\bigl\{(\Psi_{+}-\Psi_{-})/(2\Phi)\bigr\}
(\pmb{\zeta}\times\pmb{\sigma}), \label{E-js-ahe}\\
\textbf{j}^{c} &=&\sigma_{\|}\mathcal{E}\pmb{\sigma} + \sigma_{\|}\mathcal{E}\bigl\{(\Psi_{+}+\Psi_{-})/(2\Phi)\bigr\}
(\pmb{\zeta}\times\pmb{\sigma}). 
\label{E-jc-ahe}
\end{eqnarray}
As  shown in Fig. \ref{FigAHESHE}(b), for the incident charge current in the AHE configuration, the spin-up ($+$) and spin-down ($-$) electrons are scattered by skew scattering to opposite sides with the transition probability proportional to the terms $r_1 + r_2$ and $-r_1 + r_2$, respectively. As a result, the Hall current part in $\pmb{j}^c$ is proportional to $r_1$.  

The Hall resistivity is defined as $\rho_{\textrm{H}}=E_{\bot}/j_{\|}$.
Now $j_{\|}=\sigma_{\|}\mathcal{F}$ from Eq. (\ref{E-js-ahe}),
and $E_{\bot}=\mathcal{F}(\Psi_{+}-\Psi_{-})/(2\Phi)$ from Eq. (\ref{E-jc-ahe}). 
In the direction $\pmb{\zeta}\times\pmb{\sigma}$,
the spin current does not contribute to the electric field $E_{\bot}$.
With Eqs. (\ref{E-psipmphi}) and (\ref{E-a2-sigma0}),  the anomalous Hall resistivity $\rho_{\textrm{H}}$ can be obtained as 
\begin{equation}
\begin{split}
\rho_{\textrm{H}}^{} = \Lambda_1(V/N)^2(2m^5)^{1/2}B_1F_2
/(2\hbar^{4}e^2E_{F}^{1/2}),
\nonumber 
\end{split}
\end{equation}
where $B_1=(B_{+}+B_{-})/2$. 
$B_1$ involves only $r_{1}$:
\begin{eqnarray}
B_1 = \frac{1}{8\pi} && \int_0^{\pi}\int_0^{2\pi}
r_1^{} 
\bigl\{ 3F_0^2-4F_0F_1-4F_1^2\cos\theta\bigr\}\sin^3\theta d\theta d\phi. \notag\\
 \nonumber 
\end{eqnarray}
If we consider only the site-diagonal $n$ = 0 term  in $r_1^{}=r_1^{\textrm{o}}$,
$B_1=r_1^{\textrm{o}} (F_0^2-4F_0F_1/3)$, and $\rho_{\textrm{H}}^{}$  will reduce  to Kondo's result.
We remark  that the definitions of  the local  $r_1^{\textrm{o}}$ and $r_2^{\textrm{o}}$ terms were given by Kondo. 
Only the $r_1^{\textrm{o}}$ term was found to contribute to the AHE \cite{Kondo}. The effect of the $r_2^{\textrm{o}}$ term, 
on the other hand, has been hidden for about 50 years. Of course Kondo's paper was written almost a decade before Dyakonov and Perel's \cite{Dyakonov}, 
whose ideas have been explored even more recently with experimental manipulation of spin currents. 
 
The difference between the AHE and the ISHE near $T_{\rm C}$ originates from the different symmetries of the incident charge current in the AHE and incident spin current in the ISHE, as  in Fig. 1. This symmetry difference gives rise to the distinct orders of spin-spin correlation: Near $T_{\rm C}$ the temperature variation of $\rho_{\rm H}$ in the AHE is determined by a three-spin correlation $r_1$, while the $\rho_{\rm ISH}$ in the ISHE is determined by  four-spin correlations $r_{2a}$ and $r_{2b}$. The difference in order can be traced back to the different terms linear in spin orbit in Eq. (\ref{E-Ham}): For charge current the term linear in local spin $S_n$ combines with two exchange terms (for the second-order Born approximation there are three powers of the Hamiltonian)  to give a third-order term in local spins, whereas for spin current any of the last three terms, with {\it two} local spin operators, combine with two exchange terms to give a four-spin operator. In particular, the third-order correlations vanish by symmetry above $T_{\rm C}$ whereas the fourth order does not. In the limit of zero momentum transfer, i.e., $\textbf{k}-\textbf{k}^{\prime} = 0$,
$r_1$ and $r_{2a}$ (=$r_{2b}$) simplify into  first- ($\chi_1$) and second-order ($\chi_2$) non-linear uniform susceptibilities \cite{Sato}, respectively,
as shown in Fig. 2. The divergence at $T_{\rm C}$ should be smeared out in the $r_1$ and $r_{2a}$ and $r_{2b}$ due to the finite momentum transfer $\textbf{k}-\textbf{k}^{\prime}$, as for the resistance \cite{Fisher}. As a result, the different symmetries  should be reflected by a single peak below $T_{\rm C}$ in the $\rho_{\rm H}$ and two peaks of opposite sign above and below $T_{\rm C}$ in the $\rho_{\rm ISH}$. 

\begin{figure}[tbp]
\includegraphics[width = 8.5 cm]{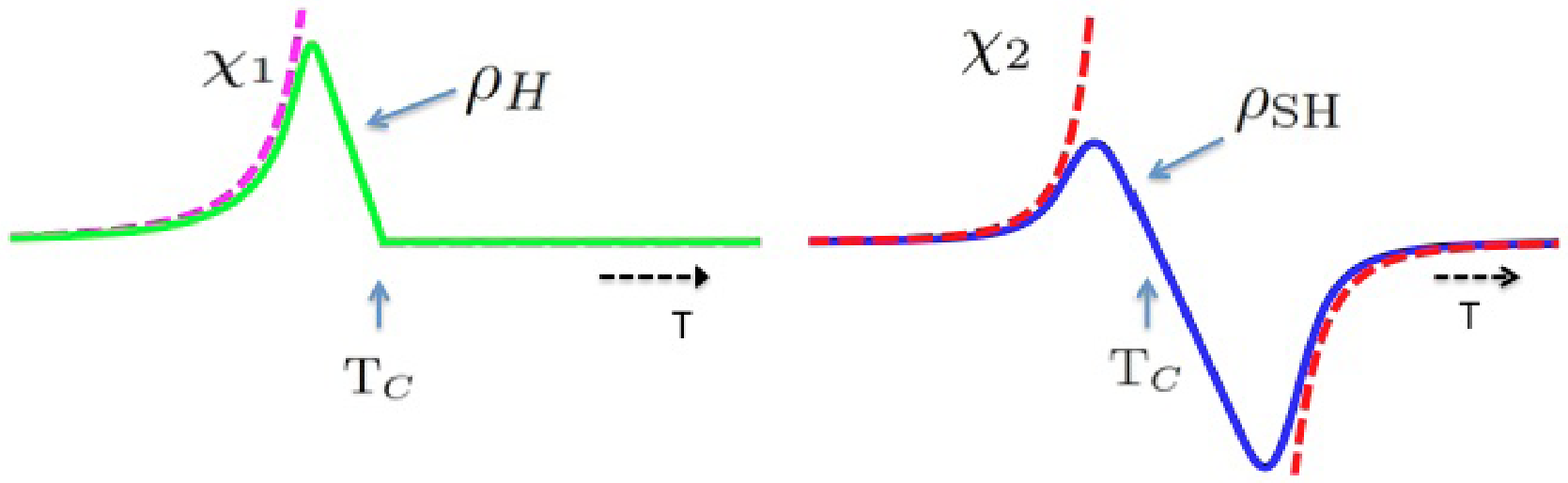}
\caption{ (Color online) Schematic behavior of the  anomalous (left) and spin Hall resistances (right) around $T_{\rm C}$. In each case the power-law divergence (dashed lines) of the corresponding  non-linear susceptibility $\chi_1$ or  $\chi_2$ is cut off by the finite Fermi surface. The different shapes (solid lines) of the anomalies reflect that the third-order correlations vanish about $T_{\rm C}$ and the fourth-order correlations change sign across $T_{\rm C}$. }
\end{figure}

\emph{Discussion}. This  theory is formulated for a pure crystalline metal  and it would  normally  be considered as an ``intrinsic" mechanism. In fact, the dependence of the Hall constant on longitudinal resistance resembles more what is classified as ``extrinsic" models for skew scattering from impurities. This is not an accident, since it is the  fluctuations of   the local moments of $d$ orbitals which skew scatter close to the Curie temperature.   Fert and co-workers \cite{Fert}, in a study of dilute alloys which are truly ``extrinsic" in the sense of including impurities, approached the transport  in a way that looks different, but is actually compatible. Rather than  finding scattering amplitudes from microscopic potentials, they replaced the $d$ orbitals by a set of phase shifts to be determined empirically.  They included   $p$- and $d$-wave phase shifts, more significant in the alloys, rather than the $s$- and $p$-scattering channels included here.  As   it is enough for skew scattering to include interference from phase shifts of different parities,  the real difference is  in the inclusion of resonant effects.  
We remark  also that expression $\rho_{\rm H}\propto r_1^{\textrm{o}}  (F_0^2-4F_0F_1/3) \Lambda_1 F_2 $, obtained in the site diagonal limit 
 is the equivalent of that of Fert {\it et al}:
$\rho_{\rm H} \propto \lambda_d\sin \delta_p \sin \left( 2\delta_d- \delta_p \right)\sin^2 \delta_d $
 with $\Lambda_1 F_2$  (coming from the scattering amplitude of the $p$-wave exchange amplitude and the local spin via the spin-orbit coupling) corresponding to the change in phase shifts for different components of $d$-wave scattering from the resonant potential $\lambda_d\sin^2 \delta_d  $.  The factor $(F_{0}^{2}-4F_{0}F_{1}/3)$ corresponds to  the term  $ \sin \delta_p \sin \left( 2\delta_d- \delta_p \right) $, 
while the former term representing interference between the $s$ wave (F$_0$) and $p$ wave (F$_1$), and the latter term 
representing interference between the $p$ wave and $d$ wave. The similarity ends here: There is no equivalent of $r_1^{\textrm{o}}$ in Fert's theory, which was formulated  for alloys at low temperatures, while here we are, as in Ref. \cite{Kondo}, interested in the collective spin fluctuations near criticality. The distinction between the anomalous and spin Hall resistivities depends on the difference between the $r_1^{}$ and the two $r_2^{}$ terms. We remark that   many-body effects can be included by use of the calculated temperature-dependent phase shifts for an isolated  quantum impurity \cite {AuFe-Gu} or in lattice models 
\cite{Coleman} for mixed-valence systems where there is a temperature-dependent  effective phase shift. 

Our theory includes skew scattering but neither side-jump nor the ``topological" lattice effects coming from anomalous velocities generated by spin-orbit terms \cite{KarplusLuttinger,Gradhand} and should thus be applicable when the conductivity is large \cite{NagaosaReview}.  
The band   theories can be extended to finite temperatures by replacing the magnetization of the band by the temperature-dependent value $\rho_{\rm H} \propto \rho_{\|}^2 \langle M \rangle$  but it is hard to see how the anomalous velocity  could produce non-monotonic behavior {\it above} $T_{\rm C}$, in contrast to our approach.  
 
\emph{Comparison with experiments}. Recently Wei {\it et al.} \cite {Otani} have performed  an inverse spin  Hall effect  experiment 
in a weak ferromagnetic alloy near $T_{\rm C}$ ,
where a dip and a peak are observed in the spin Hall resistivity $\rho_{\textrm{SH}}$ below and above the $T_{\rm C}$, respectively. We argue that our theory can  qualitatively explain the experimentally observed anomalous behaviors in $\rho_{\textrm{SH}}$ near $T_{\rm C}$. 
The  theory we provide here including correlations between spins on more than one atom seems to be necessary to explain the experiments.  From simulations \cite{Otani} on short-range Heisenberg models the on-site terms are rather smooth and only the off-site terms are non-monotonic. Thus features close to T$_{\rm C}$ may be dominated by the correlations that are off-diagonal in the site index.

\emph{Conclusions}. We have presented a theory of the inverse spin and spin Hall effects that provides an essential complement to
the classic theory of Kondo for the anomalous Hall effect, which has remained unchallenged
for 50 years. We have  found an essential difference between the anomalous  Hall effect and  the inverse spin  Hall effect near $T_{\rm C}$ in a ferromagnetic metal. Our theory can be compared to recent experimental results for the SHE in ferromagnetic alloys. 

\section*{ACKNOWLEDGEMENTS} This work was supported by 
a Grant-in Aid for Scientific Research from the
Ministry of Education, Culture, Sports, Science and Technology of
Japan.  This collaboration was supported by a REIMEI grant from the Advanced Science Research Center (ASRC) of the JAEA and T.Z. thanks the ASRC for  hospitality, as well as that of the KITP, University of California, where the manuscript was completed, with support in part by the National Science Foundation (U.S.A.) under Grant No. NSF PHY11-25915. 
The authors acknowledge  Y. Niimi, D. H. Wei,  and Y. Otani for many valuable
discussions about the experiments of SHE in ferromagnetic metals near $T_{\rm C}$ and Peter Holsdworth for providing a Monte Carlo code.


\end{document}